\documentclass[%
 reprint,
 amsmath,amssymb,
 aps,
prb,
floatfix,
]{revtex4-2}

\usepackage{braket}
\usepackage{graphicx}
\usepackage{dcolumn}
\usepackage{bm}

\usepackage{tikz}
\usepackage[caption=false]{subfig}
\usepackage{floatrow}

\usetikzlibrary{decorations.markings, calc}
\usepackage{xcolor}
\definecolor{myblue}{RGB}{68,119,170}
\definecolor{myred}{HTML}{ED6677}

\begin{document}

\preprint{APS/123-QED}

\title{Reconstructing Spin Hamiltonians of 2D Gutzwiller-Projected Wavefunctions}

\author{Lucas~Z.~Brito}
    \affiliation{Department of Physics, Harvard University, Cambridge, MA 02138}
\affiliation{Department of Physics, Brown University, Providence, RI 02912-1843}
\author{J.~B.~Marston}%
    \affiliation{Brown Theoretical Physics Center, Brown University, Providence, RI 02912-S}
    \affiliation{Department of Physics, Brown University, Providence, RI, 02912-1843, USA}

\date{\today}


\begin{abstract}
We apply the correlation matrix Hamiltonian reconstruction technique to the two-dimensional Gutzwiller-projected Fermi sea and $\pi$-flux states on finite-sized square and triangular lattices. Our results indicate no spin Hamiltonian with simple local interaction terms stabilizes such states for finite system sizes. We develop a quantitative assessment of the importance of local interactions to the stabilization of these liquid states. Lastly, we systematically assess arguments for the origin of local terms driving a Gutzwiller-projected ground state.
\end{abstract}

\maketitle


\section{Introduction}
\begin{figure}[t]
     \centering
     \vspace{1em}
     \subfloat[][]{
         \centering
         \raisebox{0.15\height}{\includegraphics[width=0.5\textwidth]{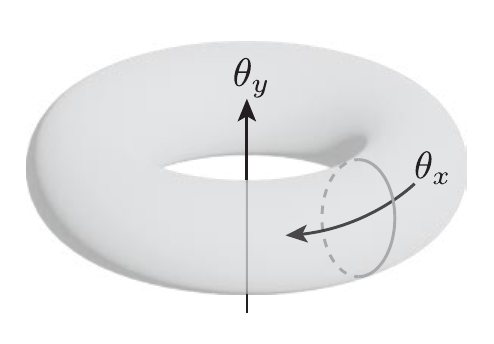}}
         \label{fig:torus-fluxes}
     }%
     \hfill
     \subfloat[][]{
         \centering
         \scalebox{1.2}{
           \begin{tikzpicture}[scale=0.9, decoration={markings, mark=at position 0.55 with {\arrow{>}}}]
    \draw[myred, thick, postaction={decorate}]  (-1, -1) -- (0, -1);
    \draw[myred, thick, postaction={decorate}] (0, -1)  -- (1, -1);
    \draw[myblue, thick] (-1,  0) -- (0, 0);
    \draw[myblue, thick]  (0,  0) -- (1, 0);
    \draw[myred, thick, postaction={decorate}] (0,  1) -- (1, 1);
    \draw[myred, thick, postaction={decorate}] (-1,  1) -- (0, 1);

    \draw[myblue, thick] (-1, -1) -- (-1, 0);
    \draw[myblue, thick] (-1,  0) -- (-1, 1);
    \draw[myblue, thick]  (0,  -1) -- (0, 0);
    \draw[myblue, thick] (0,  0) -- (0, 1);
    \draw[myblue, thick] (1,  0) -- (1, 1);
    \draw[myblue, thick] (1,  -1) -- (1, 0);

    \draw[myred, thick] (-1, -1) -- (-1.3,-1);
    \draw[myblue, thick] (-1, 0) -- (-1.3, 0);
    \draw[myred, thick] (-1, 1) -- (-1.3, 1);
    \draw[myred, thick] (1, 1) -- (1.3, 1);
    \draw[myblue, thick] (1, 0) -- (1.3, 0);
    \draw[myred, thick] (1, -1) -- (1.3, -1);

    \draw[myblue, thick] (-1, 1) -- (-1, 1.3);
    \draw[myblue, thick] (0, -1) -- (0,-1.3);
    \draw[myblue, thick] (0, 1) -- (0, 1.3);
    \draw[myblue, thick] (1, 1) -- (1, 1.3);
    \draw[myblue, thick] (1, -1) -- (1,-1.3);
    \draw[myblue, thick] (-1, -1) -- (-1,-1.3);

    \filldraw [myblue] (-1,0) circle (2pt);
    \filldraw [myblue] (0,0) circle (2pt);
    \filldraw [myblue] (1,0) circle (2pt);
    \filldraw [myblue] (-1,1) circle (2pt);
    \filldraw [myblue] (0,1) circle (2pt);
    \filldraw [myblue] (1,1) circle (2pt);
    \filldraw [myblue] (-1,-1) circle (2pt);
    \filldraw [myblue] (0,-1) circle (2pt);
    \filldraw [myblue] (1,-1) circle (2pt);

    \node at (0.5, 0.5){$\pi$}; 
    \node at (-0.5, 0.5){$\pi$}; 
    \node at (-0.5, -0.5){$-\pi$}; 
    \node at (0.5, -0.5){$-\pi$}; 

    \node (label) at (-0.5, -1.7){$e^{i\pi}$}; 
    \draw[->, line width=0.6] (label) edge [bend right=10] (-0.5, -1.2);

    \node[inner sep=1pt] (label) at (1.7, -0.25){$1$}; 
    \draw[->, line width=0.6] (label) edge [bend left=10] (1.2, -0.5);

    \draw[->, line width=0.6, gray] (-0.5,0.15) arc (270:-40:0.35);
\end{tikzpicture}
\label{fig:piflux-square}
}
     }
     \par\vspace{0.5em}
     \subfloat[][]{
         \centering
         \scalebox{1.2}{\begin{tikzpicture}[decoration={markings, mark=at position 0.55 with {\arrow{>}}}]
    \coordinate (sidea) at (0.5, 0.866);
    \coordinate (sideb) at (-0.5, 0.866);

    \coordinate (center1) at (0.5, 0.288675);
    \coordinate (center2) at (1, 0.577);

    \draw[myblue, thick]  (0, 0) -- (1, 0); 
    \draw[myblue, thick]  (1,0) -- (2,0);
    \draw[myblue, thick]  ($(sidea) + (1,0)$) -- ($(sidea)+ (2,0)$);
    \draw[myblue, thick]  ($2*(sidea) + (1,0)$) -- ($2*(sidea)+ (2,0)$);
    \draw[myblue, thick]  ($2*(sidea)$) -- ($2*(sidea)+ (1,0)$);
    \draw[myblue, thick]  (sidea) -- ($(sidea)+ (1,0)$);

    \draw[myred, thick, postaction={decorate}]  (sidea) -- ($2*(sidea)$);
    \draw[myred, thick, postaction={decorate}]  (0, 0) -- (sidea);
    \draw[myblue, thick]  ($(sidea) + (1,0)$) -- ($2*(sidea) + (1,0)$);
    \draw[myblue, thick]  (1,0) -- ($(sidea) + (1,0)$);
    \draw[myred, thick, postaction={decorate}]  (2,0) -- ($(sidea) + (2,0)$);
    \draw[myred, thick, postaction={decorate}]  ($(sidea) + (2,0)$) -- ($2*(sidea) + (2,0)$);

    \draw[myblue, thick]  (1,0) -- ($(1, 0) + (sideb)$);
    \draw[myred, thick, postaction={decorate}]  (2,0) -- ($(2, 0) + (sideb)$);
    \draw[myred, thick, postaction={decorate}]  ($(sidea) + (2,0)$) -- ($(sidea) + (sideb) + (2,0)$);
    \draw[myblue, thick]  ($(sidea) + (1,0)$) -- ($(sidea) + (sideb) + (1,0)$);

    \filldraw [myblue] (0, 0) circle (2pt);
    \filldraw [myblue] (1, 0) circle (2pt);
    \filldraw [myblue] (2, 0) circle (2pt);
    \filldraw [myblue] ($(sidea) + (0, 0)$) circle (2pt);
    \filldraw [myblue] ($(sidea) + (1, 0)$) circle (2pt);
    \filldraw [myblue] ($(sidea) + (2, 0)$) circle (2pt);
    \filldraw [myblue] ($2*(sidea) + (0, 0)$) circle (2pt);
    \filldraw [myblue] ($2*(sidea) + (1, 0)$) circle (2pt);
    \filldraw [myblue] ($2*(sidea) + (2, 0)$) circle (2pt);

    \node at (center1) {$\pi$}; 
    \node at (center2) {$0$}; 
    \node at ($(center1) + (1,0)$) {$\pi$}; 
    \node at ($(center2) + (1,0)$) {$0$}; 
    \node at ($(center1) + (1,0) + (sidea)$) {$\pi$}; 
    \node at ($(center2) + (1,0) + (sidea)$) {$0$}; 
    \node at ($(center1) + (sidea)$) {$\pi$}; 
    \node at ($(center2) + (sidea)$) {$0$}; 

    \node (label0) at (1.1,-0.5) {$1$}; 
    \draw[->, line width=0.6] (label0) edge [bend right=10] (1.5, -0.1);
    \node[inner sep=0pt] (label1) at (0,1.5) {$e^{i\pi}$}; 
    \draw[->, line width=0.6] (label1) edge [bend right=10] (0.5, 1.2);
\end{tikzpicture}}
         \label{fig:piflux-triangle}
     }
          \subfloat[][]{
          \includegraphics[width=0.4\textwidth]{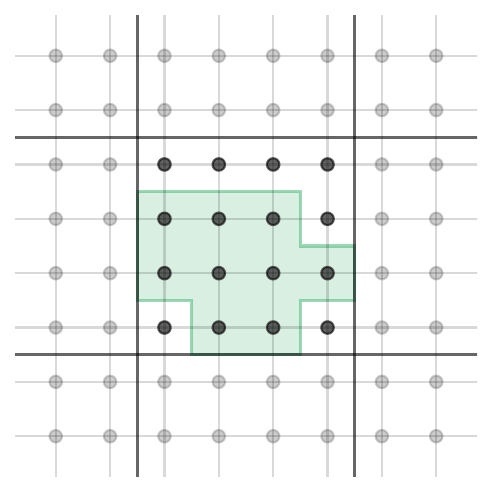}
}
     \caption{(a) The choice of nontrivial $\text{U}(1)$ holonomies around the cycles of a torus correspond to magnetic flux insertions. (b) Square $\pi$-flux ansatz flux pattern in the Landau or striped gauge. (c) Triangular $\pi$-flux ansatz flux pattern. Each pair of triangular plaquettes carries a flux $\pi$.}
     \label{fig:torus-and-piflux-tri}
\end{figure}
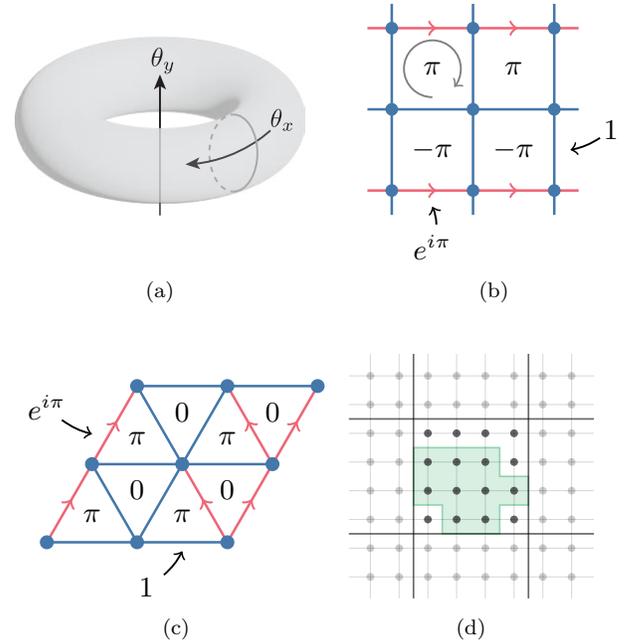

Quantum spin liquid phases have been the subject of intense research since first 
being introduced as resonating valence bond (RVB) states by Anderson in 1973 \cite{ANDERSON1973153}. These phases are, roughly speaking, characterized by a vanishing magnetic order parameter and a high degree of entanglement, preventing smooth deformation to trivial product states \cite{Savary_2017}.
The realization of such a phase is of significant interest to the condensed
matter community, as spin liquids constitute an unconventional state of matter featuring exotic phenomena such as emergent gauge fields, fractionalized excitations, and long range entanglement. 
These features also make spin liquids enticing for their potential applications, for instance as platforms for topological quantum computation or simulators of strongly interacting gauge
theories \cite{Savary_2017}.  

A systematic construction of such spin liquid ground states is achieved by the parton ansatz, wherein one fractionalizes the spin
degrees of freedom into spinful fermions and introduces static
$\text{U}(1)$ gauge degrees of fredom on the links of the lattice \cite{gros, fradkin2013field, wen, anderson}. Such a
mean-field Hamiltonian corresponds to non-interacting fermions in the presence
of a background gauge field and is exactly solvable. The gauge-inequivalent
choices of background gauge field generate a family of ground states which can be Gutzwiller-projected to the original spin Hilbert space to obtain spin liquid
states \cite{gutzwiller}. This approach can be supplemented by large-$N$ path integral techniques to argue that, for instance, the $\pi$-flux ansatz approximates the ground state energy of the nearest-neighbor square Heisenberg model \cite{Marston, PhysRevB.37.3664, PhysRevB.37.3774}.

While the unprojected mean-field ansatze are exact ground states 
of their corresponding mean-field Hamiltonians, most often projected wavefunctions are not exact ground states of any simple (e.g., nearest neighbor) spin Hamiltonian, and in fact finding a parent Hamiltonian given a projected wavefunction is a difficult task. In 1D such an exact
Hamiltonian exists for trivial background gauge configuration---the
Haldane--Shastry Hamiltonian \cite{haldane,shastry}. However, the analogous 2D 
Hamiltonian is not known for most background field choices. An
exact Hamiltonian for the two-dimensional chiral spin liquid has been proposed
\cite{Schroeter_2007}, although its form is complex and, at any rate, one would like a
general prescription for recovering exact Hamiltonians of Gutzwiller-projected
wavefunctions.

Recent years have seen a flurry of work on techniques aimed at 
recovering local Hamiltonians whose spectrum approximately or exactly includes an input wavefunction, ideally as a ground state. This program is broadly referred to as Hamiltonian design or 
Hamiltonian reconstruction \cite{qi, bairey, garrison,Hou_2020, Cao_2020, lian,  nandy2023reconstructing, turkeshi-jastrow-gutzwiller, turkeshi-parent-hamiltonians,Pakrouski2020automaticdesignof,Peschel_2003, zhang2020, zoller, jacoby, brito, kim, mei2014designlocalspinmodels, giudici, zhu}. Such techniques have seen some success in reconstructing Hamiltonians of 1D Gutzwiller projected wavefunctions and related wavefunctions \cite{turkeshi-jastrow-gutzwiller, brito}; thus, it is natural to ask whether it can be applied to their 2D analogues. One can phrase the project more precisely as follows: Much work has been directed at determining the types of interactions one must include in a spin model in order to stabilize a Gutzwiller-projected ansatz, often with the intuition that frustration must be introduced to prevent long-range ordering \cite{motrunich, mei2014designlocalspinmodels, iqbal, iqbal-pi-flux, misguich, Savary_2017}. The space of possible interactions is, however, very large. Can Hamiltonian reconstruction provide us with a systematic approach to explore the relevance of such interactions, or perhaps recover an approximate or even exact Hamiltonian for Gutzwiller-projected wavefunctions?

In this work, we study the above question with the correlation matrix reconstrution technique \cite{qi, jacoby, brito}. The correlation matrix technique begins from the simple observation that the variance of a Hamiltonian with respect to any of its eigenstates vanishes. From this statement one may define a \textit{correlation matrix} composed of correlation functions of the operators appearing in the Hamiltonian. In the appropriate conditions, this matrix is guaranteed to possess a vector in its nullspace containing the coupling constants of the Hamiltonian. In a previous work \cite{brito}, we explore what, precisely, those appropriate conditions are. We additionally demonstrate the correlation matrix exactly reconstructs the Haldane--Shastry Hamiltonian given correlators computed from the $1+1$D projected Fermi sea. This suggests the correlation matrix is amenable to reconstruction of parent Hamiltonians of $2+1$D spin liquids. In the present work, we focus on three Gutzwiller-projected wavefuntions: the projected Fermi sea on the square lattice, and the projected $\pi$-flux state on square and triangular lattices. We apply correlation matrix reconstruction on $4\times 4$ lattices, exploring the space of operators spanned by long-range spin interactions and strings of permutation operators, motivating this choice with a degenerate perturbation theory argument.

The remainder of this work is structured as follows. In Sec.~\ref{sec:models} we introduce the parton ansatze we will employ in this work and discuss choices of boundary conditions corresponding to $\text{U}(1)$ gauge field holonomies. In Sec.~\ref{sec:corr-mat} we review correlation matrix reconstruction and restate key conclusions of \cite{brito}. In Sec.~\ref{sec:operators} we discuss the space of operators one should consider when reconstructing Gutzwiller-projected wavefunctions. We motivate our choice of operator basis by perturbatively expanding an effective Hamiltonian acting on the single-occupancy sector of Hilbert space and numerically evaluate the validity of this argument. In Sec.~\ref{sec:results} we report the results of applying the correlation matrix reconstruction procedure to each of the wavefunctions introduced in Sec.~\ref{sec:models}, and in Appendix~\ref{app:operator-relevances} we suggest a technique for evaluating the relevance of each operator in the reconstruction basis, and report the results of applying this technique to our operator basis.

\section{Models}\label{sec:models}
In this work we consider three species of Gutzwiller-projected wavefunction each corresponding to a different choice of mean-field ansatz. Recall that Gutzwiller-projected wavefunctions are constructed by first diagonalizing an exactly solvable mean-field Hamiltonian 
\begin{equation}\label{eq:Hmf}
    H_\text{MF} = \sum_{\langle ij \rangle} \chi_{ij} c_i^\dagger c_j + \text{H.c.}.
\end{equation}
Here $c_i$ are parton creation an annihilation operators corresponding to the fractionalization of spin degrees of freedom, $\mathbf{S}_i = c_{i\alpha}^\dagger \boldsymbol{\sigma}_{\alpha\beta} c_{i\beta} $, and $\chi_{ij}$ are $\text{U}(1)$ background gauge link variables satisfying $\chi_{ij} = \chi^\ast_{ji}$ \cite{wen2004quantum, fradkin2013field}. This means the gauge-inequivalent background configurations are parameterized by the gauge-invariant holonomies $\chi\chi\chi\chi$, i.e., Wilson loops, or fluxes. Eq. \eqref{eq:Hmf} is an exactly solvable, quadratic Hamiltonian that possesses as groundstate a sea of free fermions we denote $\ket{\Psi_\text{MF}}$.
Different choices of $\chi_{ij}$ correspond to different choices of mean-field ansatze. 

Hamiltonians in the family \eqref{eq:Hmf} are introduced to study spin-liquid phases of spin Hamiltonians of the form 
\begin{equation}
    H = \sum_{i,j} J_{ij}\; \mathbf{S}_i\cdot \mathbf{S}_j
\end{equation}
The fractionalization of spin degrees of freedom enlarges the Hilbert space of the theory; thus, in order to recover an ansatze wavefunction in the physical Hilbert space, one \textit{Gutzwiller-projects} $\ket{\Psi_\text{MF}}$ by applying the projector $P_G = \prod_i (1-n_{i,\uparrow} n_{i,\downarrow}) $ annihilating double- and single-occupied sites \cite{gutzwiller, anderson, gros}. This technique thus allows us to systematically construct long-range resonating valence bond (RVB) spin singlet wavefunctions $P_G\ket{\Psi_{\text{MF}}}$ possessing translation invariance and lacking long-range magnetic order.

In what follows we consider three projected wavefunctions. First, the projected Fermi sea, induced by constant $\chi_{ij} = -\chi$. This wavefunction possesses the full point group symmetry of the square lattice and is an $\text{SU}(2)$ singlet. In one dimension, the analogous projected wavefunction is known to be the exact ground state of the Haldane--Shastry model \cite{haldane, shastry}; i.e., taking one of the torus side lengths to one produces a projected wavefunction which is known to posses an exact Hamiltonian. Indeed, in Ref. \cite{brito} it is shown that the Haldane--Shastry Hamiltonian is reconstructed from two-point correlators of the one dimensional projected Fermi sea via the correlation matrix. This makes the two-dimensional Fermi sea---the simplest two-dimensional extension of the ground state of the Haldane--Shastry ground state---a natural candidate for our study.

Secondly, we consider the $\pi$-flux ansatz \cite{Marston, PhysRevB.37.3774, PhysRevB.37.3664}, which is specified by 
a flux $\pi$ around all elementary plaquette, on both square and triangular lattices. In this work we utilize the `striped' gauge for the square lattice displayed in Fig.~\ref{fig:piflux-square}. Notice this gauge explicitly breaks the $\mathbb{Z}_L\times\mathbb{Z}_L$ translation symmetry to the subgroup generated by translations that preserve the flux pattern. Nevertheless the projected ground state, which is necessarily gauge invariant, possesses full translational symmetry. As with the projected Fermi sea, this wavefunction is also a $\text{SU}(2)$ singlet.

Lastly, we consider the $\pi$-flux ansatz on the triangular lattice, with flux pattern as shown in Fig. \ref{fig:piflux-triangle} \cite{Wietek_2024, iqbal-pi-flux, lu-pi-flux}. In this lattice geometry, this ansatz corresponds to zero flux through down-facing triangles and $\pi$ flux through up facing triangles. Thus the unit cell is doubled with $\pi$ flux through every unit cell. As with the square lattice $\pi$-flux state, this gauge explicitly breaks the crystalline symmetry group, which is restored by Gutzwiller projection, and likewise it is an $\text{SU}(2)$ singlet. The $\pi$-flux ansatz on triangular and square lattices is often referred to as the Dirac spin liquid (DSL) on account of its spectrum of linearly dispersing massless fermions pairing to form Dirac fermions. The triangular-lattice DSL has recently drawn attention as a potential description of the critical point of a next-nearest neighbor Heisenberg model, suggested by high overlaps with the exact ground state \cite{Wietek_2024}. One may then ask whether this state is exactly stabilized by the inclusion of longer-ranged interactions or ring exchanges. Among our results is evidence that the DSL is not the exact ground state of any such model for finite system size.

\begin{figure}[t]
     \centering
     \vspace{1em}
     \subfloat[][]{
        \includegraphics[width=0.49\textwidth]{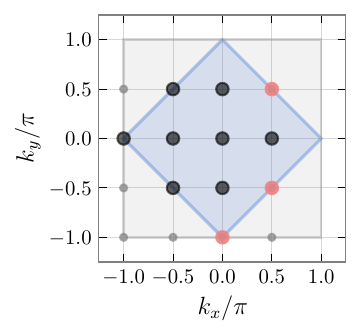}
        \label{fig:fsea-naive}
     }
     \subfloat[][]{
          \includegraphics[width=0.49\textwidth]{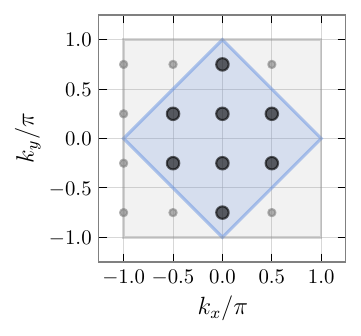}
          \label{fig:fsea_centered}
     }
     \par\vspace{0.5em}
     \subfloat[][]{
          \includegraphics[width=0.49\textwidth]{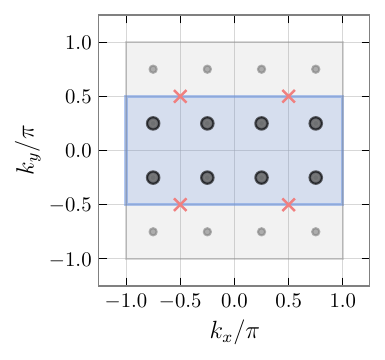}
          \label{fig:fsea-square-piflux}
     }
    \subfloat[][]{
          \includegraphics[width=0.49\textwidth]{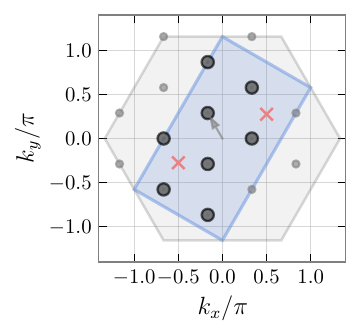}
          \label{fig:fsea-triangular-piflux}
     }
     \caption{Brillouin zones of the mean-field ansatze used in this work. (a) The projected Fermi sea with trivial (periodic) boundary conditions is degenerate. The wavevectors marked in red correspond to the degenerate choices of Fermi sea filling. (b) The projected Fermi sea with antiperiodic boundary conditions in the $y$ direction produces a non-degenerate ground state. (c) The Brillouin zone of the $\pi$-flux ansatz on the square lattice with centered boundary conditions. (d) Triangular $\pi$-flux ansatz with centered boundary conditions. The red markings indicate the Dirac points.}
     \label{fig:fseas}
\end{figure}

We close this section by addressing an important ambiguity of Gutzwiller-projected wavefunctions $\ket{P_G\Psi_\text{MF}}$ generally. Spatial manifolds with nontrivial cycles admit gauge-invariant flat connections with nontrivial holonomies, or Wilson loops. Equivalently, for a theory with gauge group $G$, one may insert $G$-defects twisting boundary conditions, i.e., one may insert magnetic fluxes through the cycles (Fig. \ref{fig:torus-fluxes}). These degrees of freedom are not `integrated out' by Gutzwiller projection and thus parameterize a family of projected wavefunctions $\ket{P_G\Psi_\text{MF}(\theta_x,\theta_y)}$ \cite{liu, wen}. In the present case, where the spatial manifold is the 2-torus and gauge group is $\text{U}(1)$, flux insertions $(\theta_x, \theta_y)$ along  have the simple physical interpretation of shifting the Fermi sea. 

It is common in the parton construction literature to treat $(\theta_x, \theta_y)$ as variational parameters that are tuned to minimize an energy functional specified by a given Hamiltonian \cite{liu, Wietek_2024}. In our case, absent a Hamiltonian, there is no obvious favored value of $(\theta_x, \theta_y)$---in fact, one may construct superpositions of projected wavefunctions with different gauge defect angles, i.e., $\int d\boldsymbol{\theta} \; \phi(\theta_x, \theta_y) \ket{P_G\Psi_{\text{MF}} (\theta_x,\theta_y)}$ for arbitrary amplitudes $\phi(\theta_x, \theta_y)$. Nonetheless, the correlation functions of the ground state, and thus reconstruction with the correlation matrix, depend on this choice. 

With this in mind, we elect to reconstruct each of the three above wavefunctions with the following choices of boundary conditions:
\begin{enumerate}
\item Trivial flux insertion, $\boldsymbol{\theta} = (0, 0)$. Note that the triangular $\pi$-flux ansatz does not support these boundary conditions. This is because at the Dirac points, the single-particle eigenstate vanishes. A simple way to see this is to note that at the Dirac points the ground state is degenerate under particle-hole transformation. However, transformation is analogous to a lattice inversion, which is explicitly broken by the flux pattern in Fig. \ref{fig:piflux-triangle} \cite{budaraju}.

\item ``Centered'' boundary conditions \cite{Wietek_2024}. In the case of the projected Fermi sea this corresponds to $\boldsymbol{\theta} = (0, \pi/L)$, i.e., antiperiodic boundary conditions in the $y$-direction. These boundary conditions are commonly chosen to eliminate the degeneracy in the Fermi sea exhibited by trivial flux insertion wavefunctions (Fig. \ref{fig:fsea-naive}). Note that this choice explicitly breaks lattice reflection symmetry. For the $\pi$-flux ansatze on square and triangular lattices, centered boundary conditions correspond to maximizing the distance between all wavevectors in the Fermi sea and the Dirac points (red markings on Figs. \ref{fig:fsea-square-piflux} and \ref{fig:fsea-triangular-piflux}). This is $(\pi/L,\pi/L)$ for the square lattice and $4\pi/6L \cdot(-1, \sqrt{3})$ for triangular lattices with $L/2$ even and  $2\pi/6L \cdot(-1, \sqrt{3})$ for $L/2$ odd.

\item Diagonalized boundary conditions. Following Ref. \cite{liu}, one can define a Gram matrix $\rho(\boldsymbol{\theta},\boldsymbol{\theta}') \equiv \braket{P_G \Psi_{\text{MF}}(\boldsymbol{\theta})\vert P_G \Psi_\text{MF}(\boldsymbol{\theta}')}$. $\rho(\boldsymbol{\theta}, \boldsymbol{\theta}')$ is numerically constructed by discretizing $\boldsymbol{\theta} \in S^1\times S^1$ into an $N\times N$ torus, where we choose $N = 8$. This Gram matrix may then be diagonalized to obtain an orthonormal basis for the subspace spanned by $\ket{P_G\Psi_\text{MF}(\boldsymbol{\theta})}$. 
\end{enumerate}

 An additional treatment of boundary conditions may be considered. It can be shown that the Berry's phase term capturing the physics of $\boldsymbol{\theta}$ is the Lagrangian of a particle on a torus pierced by uniform magnetic flux \cite{wen, liu}. This problem has been exactly solved by Haldane and Rezayi \cite{haldane-rezayi}. The theory possesses a $k$-degenerate ground state with $\phi(\boldsymbol{\theta})= e^{-kL^2\theta_x^2/4\pi}\theta_{\frac{\ell}{k}, 0}(\frac{L}{2\pi k}(\theta_y - i \theta_x) \vert ik)$, where $\theta_{a,b}(z\vert \tau)$ is the Jacobi theta function with characteristics $a,b$ and modular parameter $\tau$, and $k$ is identified as the Chern number of the Berry connection $-i\bra{P_G \Psi_{\text{MF}}(\boldsymbol{\theta})}\partial_\tau\ket{P_G \Psi_{\text{MF}} (\boldsymbol{\mathbf{\theta}})}$. In the present case, this Chern number is ill-defined. We find that for the projected Fermi sea, which is a real wavefunction, $k=0$, and in the case of the triangular and square $\pi$-flux ansatze the Chern number fails to converge as we increase the $\boldsymbol{\theta}$ discretization.

 The one dimensional projected Fermi sea $\ket{P_G\Psi_\text{MF}(\theta)}$ is parameterized by a single flux $\theta$ through the unique nontrivial cycle of the circle $S^1$. We observe that the correlation matrix reconstructs the Haldane--Shastry Hamiltonian from the projected Fermi sea for any such choice of $\theta$, i.e., $\ket{P_G\Psi_\text{MF}(\theta)}$ is the ground state of the Haldane--Shastry Hamiltonian for any $\theta$. 


\section{Correlation matrix reconstruction}\label{sec:corr-mat}
In this section we review Hamiltonian reconstruction from the correlation matrix \cite{qi, brito, jacoby}. Correlation matrix reconstruction with the simple observation that the variance of the Hamiltonian with respect to any eigenstate vanishes: 
\begin{equation}\label{eq:variance-vanishes}
    \langle H^2 \rangle - \langle H \rangle^2 = 0
\end{equation}
One assumes $H$ is a local, Hamiltonian, i.e., it may be written as sum of terms:
\begin{equation}\label{eq:H-local}
    H = \sum_i^N J_i O_i
\end{equation}
where $J_i\in \mathbb{R}$ are coupling constants and $O_i$ are Hermitian operators. We assume the Hamiltonian can be expressed as a finite sum, although $O_i$ may contain a sum of infinitely many operators, such as a sum over sites in translationally invariant systems; thus such Hamiltonians may model infinite systems. Typically $O_i$ are assumed to be local, i.e., they act non-trivially on a finite number of spatially adjacent sites, although in principle the only requirement of $O_i$ is that correlation functions $\langle O_i O_j\rangle$ are computable. Eq. \eqref{eq:variance-vanishes} may then be written 
\begin{equation}
    \vec{\gamma}_H^\text{T} \mathcal{M} \vec{\gamma}_H = 0 
\end{equation}
where $\mathcal{M}$ is the correlation matrix and $\vec{\gamma}_\text{H}$ is a vector of coupling constants:
\begin{gather}
    \mathcal{M}_{ij} = \frac{1}{2}\langle \left\{ O_i, O_j\right\} \rangle - \langle O_i\rangle \langle O_j\rangle\\ 
    \vec{\gamma}_H = (J_1, J_2, \cdots, J_N)
\end{gather}
Here the anticommutator $\left\{\cdot,\cdot\right\}$ is applied to produce a symmetric $\mathcal{M}$. Crucially, an immediate consequence is that $\vec{\gamma}_H$ is in the nullspace of $M$, i.e., $\gamma_H \in \ker{\mathcal{M}}$. As such, if $\dim \ker \mathcal{M} = 1$, $H$ may be read off (up to an overall positive multiplicative constant) by diagonalizing $\mathcal{M}$ and inspecting the nullspace. More generally, given a wavefunction $\ket{\Psi}$, which we refer to as the input wavefunction, we may define a correlation matrix 
\begin{equation}
        \mathcal{M}_{ij}^\Psi = \frac{1}{2}\langle \left\{ O_i, O_j\right\} \rangle_\Psi - \langle O_i\rangle_\Psi \langle O_j\rangle_\Psi.
\end{equation}
We denote $\lambda^{(k)}$ the eigenvalues of $\mathcal{M}^\Psi$, $\vec{\gamma}^{(k)} = (\gamma_1^{(k)}, \gamma_2^{(k)}, \cdots \gamma_N^{(k)})$ the eigenvectors, and $\Gamma^{(k)} = \sum_i \gamma_i^{(k)} O_i$ the operators constructed from the spectrum of $\mathcal{M}^\Psi$. Then $\Gamma^{(k)}$ may be interpreted as a set of operators with mutually vanishing covariance, and $\lambda^{(k)}$ are the variances of these operators with respect to the input wavefunction. In the remainder of this paper we omit the subscript $\Psi$ from expectation values $\langle\cdots\rangle_\Psi$ and from the correlation matrix $\mathcal{M}^\Psi$; expectation values will always be taken with respect to one of the Gutzwiller wavefunctions listed in the preceding section.

Thus, in principle, a Hamiltonian may be reconstructed from a single eigenstate $\ket{\Psi}$ via the following procedure: (1) choose an appropriate \textit{reconstruction operator basis} $\left\{O_i\right\}$. (2) Compute the connected correlators with respect to the input wavefunction $\ket{\Psi}$ of each pair of operators in the reconstruction basis in order to build $\mathcal{M}$. (3) Diagonalize $\mathcal{M}$ and read the Hamiltonian off the nullspace; i.e., $\Gamma^{(0)} \propto H$.

In Ref. \cite{brito}, we discuss several factors complicating the procedure outlined above. Firstly, if conserved operators $Q$ are in the span of the reconstruction basis, the fact that $\ket{\Psi}$ is a simultaneous eigenstate of $H$ and $Q$ leads to $\dim\ker\mathcal{M}>1$. Thus the Hamiltonian cannot be uniquely read off and lies in the space of operators spanned by $H$ and $Q$. 

Secondly, successful reconstruction depends sensitively on the choice of operator basis, and what constitutes an ``appropriate'' basis is subtle. In many settings, one might not have \textit{a priori} knowledge of the correct reconstruction basis, or might not have access to a full set of correlation functions. In the present case, the chief complication is the lack of a preferred reconstruction basis; we discuss our strategy for selecting such a basis in the following section. 

Note that even if the reconstruction basis coincides with the Hamiltonian's operator basis, if the operators in those operators are not linearly independent at the level of correlation functions, additional zeros will appear in the operator basis and complicate reconstruction. This is because those operators can be summed to obtain the zero operator, which trivially has vanishing variance with respect to any state. In this case, one must modify the reconstruction basis by picking one representative out of the set of linearly dependent operators. The Hamiltonian will still appear in the nullspace as this amounts to regrouping terms in Eq. \ref{eq:H-local}.

In Ref. \cite{brito} it is argued that reconstruction from an incomplete operator basis produces an ``approximate'' Hamiltonian with variance---i.e., $\mathcal{M}$ eigenvalue---approximately scaling as the square of the largest missing coupling constant. This means that, when reconstructing with an incomplete basis, the lowest eigenvalue (that is not associated with a conserved operator $Q_i$) may be used as a proxy for the ``distance'' to an exact Hamiltonian. This makes reconstruction from the correlation matrix a form of variational technique which outputs the minimum-variance operator in the span of the reconstruction basis with respect to a given wavefunction.

We close by noting a significant shortcoming of the correlation matrix technique. 
While the correlation matrix produces an operator which is guaranteed to have minimum variance with respect to the input wavefunction, there are no constraints on the expectation value with respect to that operator. In other words, successful reconstruction produces an operator which possesses $\ket{\Psi}$ as an eigenstate, but not necessarily a ground state. That $\ket{\Psi}$ is a ground state of $H$ is rather a loose expectation based on the properties of the ground state wavefunction---e.g., in our case, there are no spinon excitations above the Fermi surface, etc. 
 
\section{Space of operators}\label{sec:operators}
As discussed above, the chief complication in reconstructing Hamiltonians with the correlation matrix is the choice reconstruction basis. In the present case, the electronic degrees of freedom are frozen out and the space of operators is restricted to the span of products of Pauli operators. Further restrictions on this space are imposed by symmetries of Gutzwiller-projected wavefunctions. Firstly, the on-site spin-rotational symmetry restricts us to rotationally-invariant combinations of spin operators, that is, scalars formed out of $\mathbf{S}_i$. 

Secondly, the translational invariance of $P_G\ket{\Psi}$ permits us to choose a basis of spatially-averaged operators. Materially, this means that a basis of composed of the $(L^2)^n$ possible (spin-invariant) products of $n$ local operators may be reduced to a basis of $(L^2)^{n-1}$ products of $n$ operators where the first operator in the product is spatially-averaged, i.e., summed over all lattice sites. Both these simplifications are consequences of the more general requirement that if the wavefunction enjoys a symmetry group $G$, the elements of the reconstruction basis must be symmetrized with respect to $G$.  

Lastly, operators may act at separations identified with each other under the periodic boundary conditions. In this case one must pick a representative of each set of such identified operators. In the case of the 2-torus, we identify the unique two-body operators as those bounded by the green region in Fig. \ref{fig:torus-and-piflux-tri}d.

Despite the significant simplification the above considerations induce, the size of the space of spin-invariant, translationally-averaged products of spin operators remains formidable. One then asks if there is a systematic way to explore this space of operators. Insofar as we are concerned with Gutzwiller-projected wavefunctions, there are some common choices of operators historically proposed to drive spin-liquid phases. For instance, long-range spin-exchange interactions of the form $\mathbf{S}_i\cdot\mathbf{S}_j$ are known to stabilize the Gutzwiller-projected Fermi sea in the case of the Haldane--Shastry model. In two-dimensions, so-called ring exchange terms have also been the subject of intense study. Nevertheless, this is a small corner of the space of operators, and we seek an organizing principle that may inform our choice of operator basis.

\begin{figure}
    \centering
    \includegraphics[width=1\linewidth]{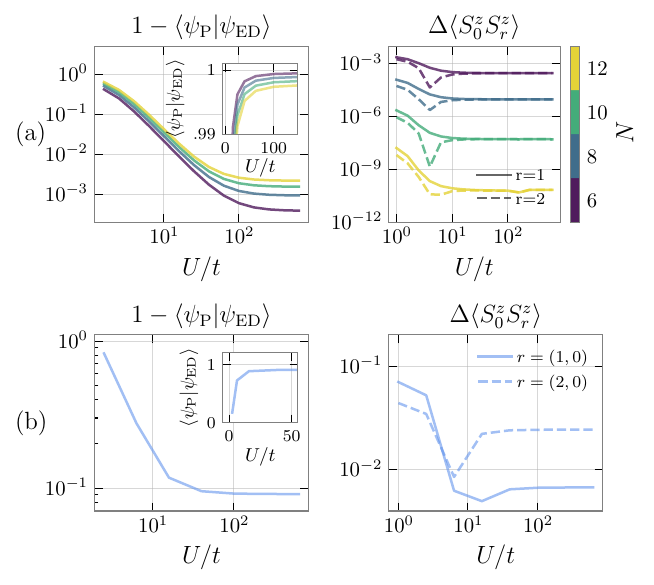}
     \caption{Wavefunction overlaps and difference in two-point correlators of Gutzwiller-projected ground state and exact ground state as a function of single-occupancy constraint strength. (a) Tight-binding model in $d=1$ with $N=10$ sites, whose projected ground state is the exact ground state of the Haldane--Shastry model. (b) $\pi$-flux state in $d=2$ on $4\times 4$ square lattice. Here $\Delta\langle S_0^z S_r^z\rangle \equiv \langle S_0^z S_r^z\rangle_\text{ED} - \langle S_0^z S_r^z\rangle_\text{P}$.}
    \label{fig:pert}
\end{figure}

In this work we adopt the strategy employed in Ref. \cite{lu}; specifically, we begin with a mean field Hamiltonian of \eqref{eq:Hmf} and introduce the following `soft-constraint' punishing double occupancy and vacant sites: 
\begin{equation} \label{eq:h-int}
    H = H_\text{MF} + U \sum_i (n-1)^2.
\end{equation}
One then tunes to a large $U\gg t$, thus driving the system to a Mott-insulating phase that is expected to at least approximate $P_G\ket{\Psi}$. When $U$ is very large, we expect the low-energy dynamics to be effectively governed by a spin Hamiltonian corresponding to the single-occupancy sector; in this case the effective Hamiltonian can be derived by means of degenerate perturbation theory with degenerate subspace corresponding to the single-occupancy sector. Specifically, writing $H = H_{\text{MF}} + H_\text{int}$, we find 
\begin{align}\label{eq:pert-h}
    H_\text{eff} &= PH_\text{MF} \left(
    1 + \Gamma H_\text{MF}
    \right)^{-1} P  \\ 
    &= PH_\text{MF}P - P H_\text{MF} \Gamma P_{\text{MF}} P \nonumber\\ 
    &\qquad \qquad + PH_\text{MF} \Gamma H_{\text{MF}} \Gamma H_\text{MF} P+\cdots, \nonumber 
\end{align}
where $P$ projects to the single-occupancy sector, $Q = 1-P$ projects to the remainder of the Hilbert space, and $\Gamma = QH_{\text{int}}^{-1}Q$. In other words, the action of the interacting Hamiltonian \eqref{eq:h-int} may be organized as a series of electron hoppings along strings of neighboring sites. Specifically, the order $\chi^n/U^{n-1}$ hopping terms in the above expansion corresponds to $n$ fermion hoppings, each followed by a projection to the doubly-occupied subspace and the accumulation of a $1/U$ factor, followed by a final projection to the singly-occupied subspace. In the single-occupied subspace these term acts as strings of nearest-neighbor permutation operators
\begin{equation}
    P_\gamma = \left(\prod_{\langle i, j \rangle \in \gamma} P_{ij}\right) + \text{H.c.}
\end{equation}
where $P_{ij} = 2\mathbf{S}_i\cdot\mathbf{S}_j + \frac{1}{2}$
\footnote{Products of non-contiguous permutation operators are allowed but expand to sums of spatially contiguous permutations.}.

In order to assess the validity of the above argument, we use exact diagonalization to directly compute the ground state of the above model for a range of $U$ values, and compare this family of ground states to the analogous Gutzwiller-projected wavefunction \cite{xdiag}. Specifically, we project the family of exact ground states to the single-occupancy sector and quantify their similarity to the Gutzwiller-projeted states by comparing two-point spin correlators and taking wavefunction overlaps. Results are shown in Fig. \ref{fig:pert} for the one-dimensional projected Fermi sea, the two-dimensional projected Fermi sea, and the $\pi$-flux state. 

Our expectation is that if this effective Hamiltonian exactly stabilizes the Gutzwiller-projected wavefunction, the family of exact ground states should approach $\ket{P_G \Psi}$ in overlap and correlation functions in the large $U$ limit. Results 
are shown in Fig. \ref{fig:pert}a for the projected Fermi sea in $d=1$ for a $N=10$ chain, and Fig. \ref{fig:pert}b for the $pi$-flux state in $d=2$ for a $4\times 4$ square lattice. We find that as $U$ is increased, the wavefunctions' overlaps plateau to a value close to but less than unity, and the difference in correlators remains nonzero, suggesting that a strong soft single-occupancy constraint stabilizes a projected ground state approximately but not exactly; conversely, we conclude that the effective Hamiltonian acting on the single-occupied sector only approximately captures the fluctuations experienced by Gutzwiller-projected states. However, as Fig.~\ref{fig:pert}a indicates, it is possible that in the infinite system limit the two ground states agree.

One may ask if the amplitudes contaminating the exactly-diagonalized ground state in the $U\to \infty$ limit lie outside of the single-occupied sector; i.e., whether applying $P_G$ to $\ket{\Psi_\text{ED}(U)}$ recovers the Gutzwiller-projected state and the soft single-occupancy constraint agrees with the hard constraint if the wavefunctions are restricted to the the single-occupancy sector. Indeed, the degenerate perturbation theory derives an effective Hamiltonian restricted to the single-occupied sector. We find even in this case the wavefunctions do not agree: projecting $\ket{\Psi_{\text{ED}}}$ for $U/t \sim 10^3$ to the single-occupied sector produces an overlap of $1-\epsilon$ with $\epsilon \approx 0.0913$ for the $d=2$ $\pi$-flux state and $\epsilon\approx 0.0015$ for the $d=1$ projected Fermi sea, signaling that the contamination is not orthogonal to the space of spin wavefunctions.

We nevertheless include the operators generated by this perturbative expansion in our reconstruction basis with hopes of systematically studying the relevance of these terms to the stabilization of Gutzwiller-projected ground states. Our reconstruction basis thus consists of
\begin{itemize}
    \item Spatially-averaged long-range Heisenberg interactions: 
    \begin{equation}\label{eq:ss-basis}
        \mathcal{S} = \left\{ \frac{1}{N}\sum_{i}\mathbf{S}_i\cdot\mathbf{S}_{i+\Delta r} :  \Delta r \in \Lambda' \right\}
    \end{equation}
    Here $\Lambda'$ is the set of all lattice separations that does not include sites identified on account of the periodic boundary conditions; these are the sites bounded by the green region in Fig. \ref{fig:torus-and-piflux-tri}d. This is simply the two-dimensional analogue of the torus-resolved basis of Ref. \cite{brito}.

    The span of this basis contains the total spin, squared, $\left(\sum_i\mathbf{S}_i\right)\cdot \left(\sum_i\mathbf{S}_i\right)$. Since the wavefunctions above are $\text{SU}(2)$ singlets, this is a conserved operator, i.e., it appears in the nullspace of the correlation matrix. Therefore, when inspecting the spectrum of $\mathcal{M}$, one must track the lowest nonzero eigenvalue. We henceforth refer to the lowest eigenvalue not corresponding to this conserved quantity as simply ``the lowest eigenvalue,'' denoted $\lambda_0$.
    
    \item Spatially-averaged open and closed strings of spin permutation operators:
    \begin{equation}
    \mathcal{P}_k = 
        \left\{\frac{1}{N} \sum_i P_{\gamma^k (i)} : \gamma^k(i)= \substack{\text{string of }k\text{ links}\\\text{starting at } i}\right\},
    \end{equation}
    Including all possible paths---even on $4\times 4$ lattices---is computationally unfeasible and we restrict to paths with $2\leq k\leq 4$ for square lattices and $2\leq k\leq 3$ for triangular lattices. 
    Note this set includes ring exchange interactions as a special case \cite{Thouless, motrunich}. 
    Further, expanding and applying Pauli algebra identities reveals each path generates Heisenberg exchange terms between the endpoints of the paths, in addition to other terms involving cross products of $\mathbf{S}_i$ operators. Note that paths of length $k=2$ reduce to sums over operators included in \eqref{eq:ss-basis}. 
\end{itemize}

In other words, our reconstruction basis is $\mathcal{O} \equiv \mathcal{S} \cup \mathcal{P}_k$ and we
explore the space of Hamiltonians of the form 
\begin{equation}
    H(\left\{J, K\right\})
        = \sum_{i, j} J(|i - j|) \mathbf{S}_i\cdot\mathbf{S}_j 
            + \sum_{\gamma^k \in \Gamma} K(\gamma^k) \sum_i P_{\gamma^k(i)} 
\end{equation}
where $\Gamma$ consists of all paths (up to lattice translation) of lengths $k=2,3,4$ for square lattices and $k=2,3$ for triangular lattices. Note that $J(|i-j|)$ depends only on separation and $K(\gamma^k)$ depends only on the shape of the path and not the starting point; that is, this is a fully translationally invariant Hamiltonian. 

We mention that, in addition to the \textit{a priori} motivation offered by the above perturbative argument, the inclusion of $\mathcal{P}_k$ in the reconstruction basis provides a source of frustration. It is broadly expected that a stable spin liquid phase requires some source of frustration inhibiting conventional (e.g., N\'eel or ferromagnetic) ordering---for example, geometric frustration on triangular lattices \cite{Savary_2017}. In this case, the sign of $P_\gamma$ alternates depending on the parity of the length of $\gamma$, introducing competition between ferromagnetic and antiferromagnetic ordering. For this reason so-called ring operators on plaquettes or groups of adjacent plaquettes have been a popular choice of term for candidate spin liquid Hamiltonians \cite{motrunich, Savary_2017, Thouless}.

\section{Results}\label{sec:results}

\begin{table}[t]
\caption{\label{tab:results}%
Lowest eigenvalue of the correlation matrix, $\lambda_0 = \langle H_{\text{app}}^2 \rangle - \langle H_{\text{app}}\rangle^2 $, ground state energy of $H_\text{app}$, and $\langle H_\text{app}\rangle = \bra{P_G \Psi_\text{MF}} H_\text{app} \ket{P_G \Psi_{\text{MF}}}$.
}
\begin{ruledtabular}
\begin{tabular}{lccccc}
    &BC& $ \lambda_0 $& $\alpha$ &$ E_0 $&$ \langle H_{\text{app}}\rangle  $ \\
\colrule
Fermi sea & Triv. & $5.133\mathrm{e}{-5}$ & $-0.10$ & $-0.3375$ & $+0.1291$ \\
	 & Cent. & $1.060\mathrm{e}{-5}$ & $+0.59$ & $-0.3611$ & $-0.3454$ \\
	 & Diag. & $8.012\mathrm{e}{-5}$ & $-0.34$ & $-0.5717$ & $+0.2839$ \\
Sq. $\pi$-flux & Triv. & $2.046\mathrm{e}{-4}$ & $+0.21$ & $-0.5306$ & $+0.2375$ \\
	 & Cent. & $2.563\mathrm{e}{-6}$ & $+0.65$ & $-0.3011$ & $-0.0659$ \\
	 & Diag. & $9.479\mathrm{e}{-6}$ & $+0.60$ & $-0.3823$ & $-0.3625$ \\
Tri. $\pi$-flux & Diag. & $1.289\mathrm{e}{-5}$ & $+2.69$ & $-0.4946$ & $-0.4637$ \\
	 & Cent. & $6.495\mathrm{e}{-5}$ & $+0.49$ & $-0.1713$ & $-0.0707$ \\
\end{tabular}
\end{ruledtabular}
\end{table}

As mentioned in Sec. \ref{sec:corr-mat}, one of the complicating factors of reconstruction with the correlation matrix is the presence of zero-eigenvalues enlarging the nullspace of $ \mathcal{M} $. Such zeros can arise from conserved quantities built of out the reconstruction basis, or out of operators that are linearly dependent. In the present case, the fact that all of the wavefunctions we reconstruct from are spin singlets ensures that the nullspace is at least one dimensional, corresponding to the total spin. However, addition zero-eigenvalues appear due to the fact that the reconstruction basis $ \mathcal{O} $ is not linearly independent; in particular, certain $\mathcal{P}_k$ operators decompose to non-obvious sums of other operators in the basis. We account for this by searching for linear combinations of subsets of operators in the reconstruction basis that produce a vector in the nullspace of the matrix. Out of these linear combinations, only the conserved quantity has non-trivially vanishing variance.

Our results are reported in Table \ref{tab:results}. We remind the reader that the lowest eigenvalue of the correlation matrix $\lambda_0$ (that does not correspond to the conserved quantity) should be interpreted as the variance of an approximate Hamiltonian $H_\text{app}$ with coefficients determined by the corresponding eigenvector. In particular, we find that there is no operator other than $S_{\text{tot}}$ in the space spanned by the reconstruction basis $ \mathcal{O} $ with vanishing variance with respect to the projected Fermi sea or $\pi$-flux state on square and triangular lattices. That is, these states possess no exact Hamiltonian with long-ranged spin exchange interactions and string permutation interactions of up to length $k=4$ for square lattices and $k=3$ for triangular lattices. 

Contributions to a candidate Hamiltonian from the conserved quantity $S_\text{tot}$ are not detected by the correlation matrix, as $\langle S_\text{tot}^2\rangle - \langle S_\text{tot}\rangle^2 =0$. Therefore Hamiltonians of the form $H_\alpha = \sum_i \gamma^{(0)}_i O_i + \alpha S_\text{tot}$, for $\alpha$ in the same energy scale as $\gamma_i$ \footnote{When $|\alpha| \ll \gamma_i$, $S_\text{tot}$ dominates the Hamiltonian $H_\alpha$ and the input wavefunction is trivially an eigenstate.}, comprise a family of approximate Hamiltonians with variance equal to $\lambda_0$. 
Further, as noted in Sec. \ref{sec:corr-mat}, the correlation matrix tracks the variance of operators in the basis, but not expectation values, i.e., there is no guarantee that the approximate Hamiltonians reconstructed possess the projected wavefunctions as approximate \textit{ground states}, only that they are approximate \textit{eigenstates}. Thus, we set $\alpha$ for each of the input wavefunctions by minimizing the difference between the exact-diagonalization ground state energy of $H_\alpha$ and the expectation $\langle H_\alpha\rangle = \bra{P_G\Psi_\text{MF}} H_\alpha\ket{P_G\Psi_{\text{MF}}}$. We refer to the approximate Hamiltonian for such an optimal $\alpha_\text{min}$ as $H_\text{app}\equiv H_{\alpha_\text{min}}$.  

\begin{figure}
    \centering
    \includegraphics[width=1\linewidth]{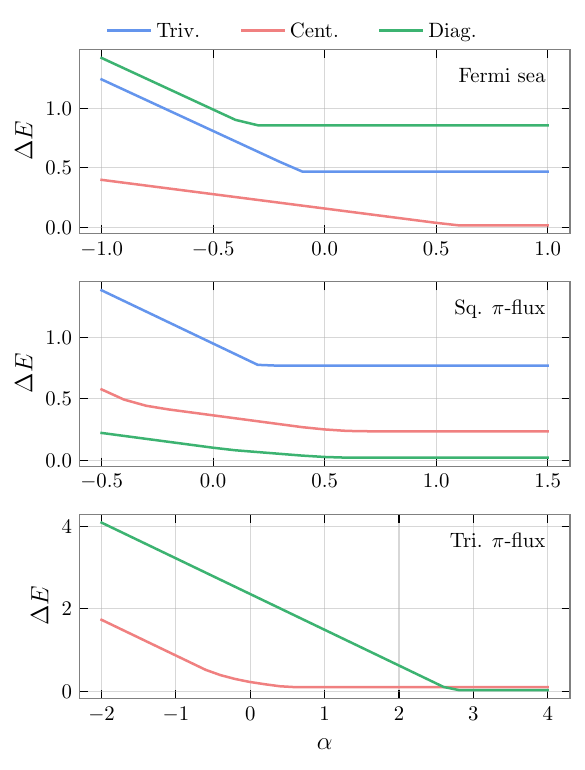}
    \caption{Difference $\Delta E \equiv \langle H_\alpha\rangle - E_0$ between exact-diagonalization ground state energy $E_0$ and expectation value $\langle H_\alpha \rangle$ of approximate Hamiltonian $H_\alpha = \alpha S_\text{tot} + \sum_i \gamma_iO_i$.}
    \label{fig:enter-label}
\end{figure}

We note that at $\alpha>\alpha_\text{min}$, the difference $E_0 - \langle H_\alpha\rangle $ remains constant. More precisely, the ground state energy of $H_\alpha$ is linear in $+\alpha$ until a critical value $\alpha_\text{min}$. At $\alpha >\alpha_\text{min}$, $S_\text{tot}$ favors antiferromagnetic ordering and the ground state energy is proportional to $-\alpha$. As the $S_\text{tot}$ term in $\langle H_\alpha\rangle$ is a constant proportional to $\alpha$, $E_0 - \langle H_\alpha\rangle$ is also constant.

While the correlation matrix does not yield an exact Hamiltonian, it may still
be leveraged to study the relevance of particular operator to the stabilization
of the input wavefunction. Specifically, one may use $\lambda_0$ to measure how
much each operator in the reconstruction basis contributes to minimizing the
variance. We perform such analysis in Appendix
\ref{app:operator-relevances}. 

\section{Conclusion}\label{sec:conclusion}
In this work we have explored Hamiltonian reconstruction of Gutzwiller-projected wavefunctions via the correlation matrix technique. We have found that on $4\times4$ lattice, the projected Fermi sea and square and triangular $\pi$-flux ansatze do not possess an exact Hamiltonian in the subspace spanned by long-range spin exchange interactions and spin permutations along finite strings. We therefore conclude that these interactions are insufficient to stabilize these projected states, meaning that, broadly speaking, a high degree of frustration from interactions. This suggests future work on parton spin liquids might turn to, say, higher-order terms in the perturbation expansion of Eq.~\ref{eq:pert-h}, or products of non-contiguous operators respecting translation and $\text{SU}(2)$ symmetry.

We also report that, interestingly, the success of the reconstruction procedure depends on the choice of boundary conditions of the projected wavefunctions, that is, the $\text{U}(1)$ holonomies. Specifically, we find that such choices have an effect on the proximity of the ground state energy of the approximate reconstructed Hamiltonian and the energy with respect to the input wavefunction (Table~\ref{tab:results}). This suggests that there is a subtle relationship between the stability of these states and the choice of $\text{U}(1)$ fluxes, and future work might explore this relationship further.

Lastly, we presented evidence that effective Hamiltonians derived from a soft occupancy constraint acting on the single-occupied sector do not exactly stabilize these states either. However, we also found evidence that, in one dimension, such an effective Hamiltonian could stabilize the projected Fermi sea in the infinite system limit. One direction for future work thus might involve exploring this procedure for larger systems. Our study was limited to $4\times 4$ lattices by the computational demand of constructing Gutzwiller-projected wavefunctions. We note that in principle it is not necessary to compute the wavefunction, only to evaluate expectation values of correlators of operators in the reconstruction basis. On that note, we explored employing variational Monte Carlo (VMC) techniques on lattices larger than $6\times 6$, and found that Monte Carlo fluctuations are too large to perform reliable reconstruction \footnote{In effect Monte Carlo fluctuations act as an additive error $\mathcal{M}+ \epsilon \mathcal{M}'$ to the correlation matrix. See Sec. 3 of \cite{qi} for a discussion of this type of error.}. Thus, to study larger systems, it will be necessary to expore alternative numerical techniques for evaluating expectation values of Gutzwiller-projected wavefunctions, e.g., tensor network methods \cite{Jin_2020, li-tensor-network-gaussian-fpeps}. 

Lastly, we comment that there are analogues of the one-dimensional Haldane--Shastry model that one could use as a benchmark for correlation matrix reconstruction in two dimensions. For instance, an exact Hamiltonian for the chiral spin liquid has been suggested in \cite{Schroeter_2007}; although this Hamiltonian contains a vast number of terms, prohibiting exact reconstruction, it is in principle possible to incrementally include operators in a reconstruction basis and attempt to recover the functional form of its couplings in the spirit of \cite{brito}. Alternatively, one might attempt reconstruction of the Kitaev honeycomb model, which posseses an exactly solved ground state \cite{Kitaev_2006}.

\begin{acknowledgments}
L.Z.B. acknowledges support from the Barry Goldwater Scholarship and Excellence in Education Foundation during part of this work. The authors thank Stephen Carr, Dan E. Parker, Alex Jacoby, Ashvin Vishwanath, Eslam Khalaf, and Aaron Hui for useful discussions.
\end{acknowledgments}

\appendix
\section{Permutation operators}\label{app:operator-relevances}
While the operator basis $\mathcal{O}$ does not generate a Hamiltonian that
possesses any of the projected wavefunctions considered in this study as an
eigenstate, one may nonetheless ask how much operator in the basis contributes
to the stabilization of the wavefunction. Specifically, one may ask how much
each operator decreases the lowest eigenvalue of the correlation matrix
$\lambda_0$. Let $ \lambda^{\mathcal{B}}_0 $ be the lowest eigenvalue of the
correlation matrix $ \mathcal{M}^{\mathcal{B}} $ constructed from a
reconstruction basis $ \mathcal{B} $. One may consider a truncated reference
reconstruction basis $\mathcal{O}^{\emptyset}$ and correspondingly a truncated reference
correlation matrix $ \mathcal{M}^\emptyset $, which yields a lowest eigenvalue
$\lambda^\emptyset_0$. Then, one adds an operator $ O $ to the basis, constructs
a new matrix $ \mathcal{M}^{O}$ with lowest eigenvalue $ \lambda_0^O $ out of
this enlarged basis, and measures $ \lambda_0^O - \lambda_0^\emptyset $. 

There are many choices of reference basis $ \mathcal{O}^{\emptyset} $. Choosing such a basis and studying the inclusion of additional operators then amounts to asking: starting from the interactions in $\mathcal{O}^\emptyset$, how do additional interactions bring the Hamiltonian closer to possessing the input wavefunction as an exact eigenstate? 

Here we choose $\mathcal{O}^\emptyset = \mathcal{S}$, that is, we
choose the basis of long-range exchange interactions. This is precisely the
reconstruction basis that recovers the Haldane--Shastry Hamiltonian in one
dimension. Thus, we are asking: starting from the two-dimensional analogue of
the Haldane--Shastry Hamiltonian---that is, a long-ranged Heisenberg model---
what is the relevance of each string permutation operator in stabilizing each projected wavefunction.

\renewcommand{\thefigure}{\arabic{figure}(a)}
\begin{figure*}[t]
    \includegraphics[width=0.9\textwidth]{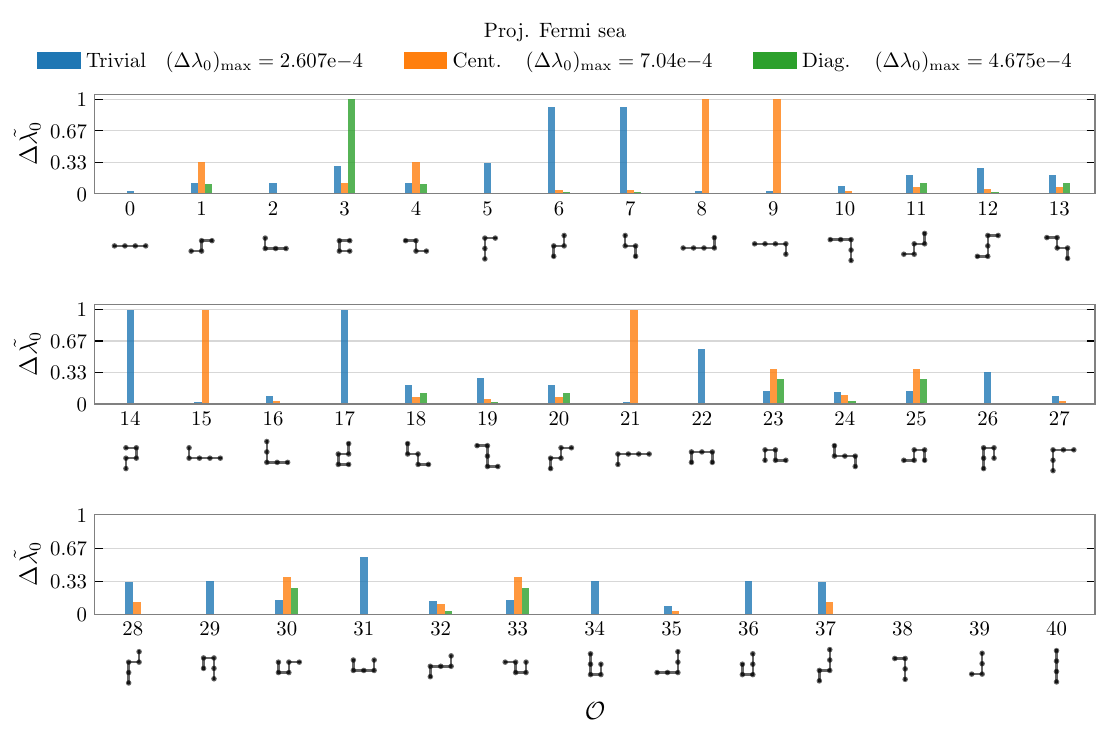}
    \caption{Normalized eigenvalue decreases for each unique permutation operator.}
    \label{fig:fsea-perms-evals}
\end{figure*}
\renewcommand{\thefigure}{\arabic{figure}(b)}

\begin{figure*}[t]
    \ContinuedFloat
    \includegraphics[width=0.9\textwidth]{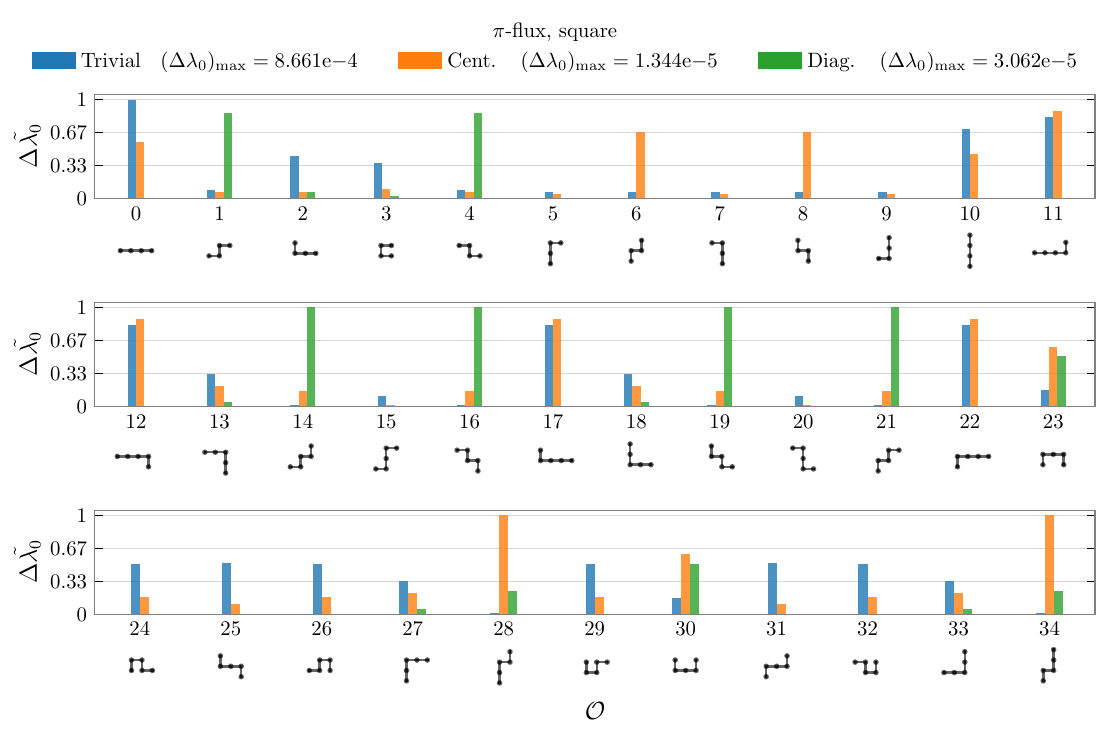}
    \caption{Normalized eigenvalue decreases for each unique permutation operator.}
    \label{fig:piflux-sq-perms-evals}
\end{figure*}
\renewcommand{\thefigure}{\arabic{figure}(c)}

\begin{figure*}[t]
    \ContinuedFloat
    \includegraphics[width=0.9\textwidth]{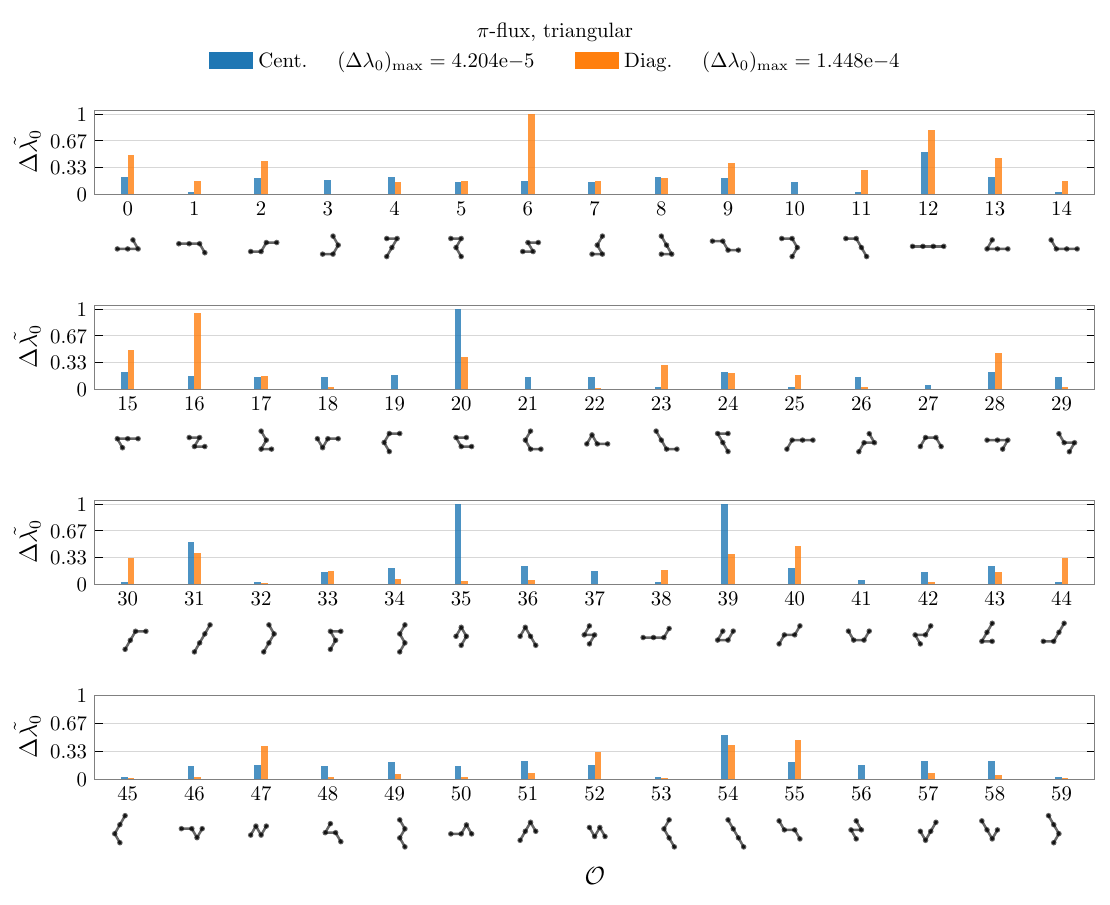}
    \caption{Normalized eigenvalue decreases for each unique permutation operator.}
    \label{fig:piflux-tri-perms-evals}
\end{figure*}

Results are displayed in Figs.~\ref{fig:fsea-perms-evals}, \ref{fig:piflux-sq-perms-evals}, and \ref{fig:piflux-tri-perms-evals}. Here $\Delta \tilde{\lambda}_0 \equiv \Delta\lambda_0 / (\Delta\lambda_0)_\text{max} $ where $\Delta\lambda_0 = \lambda_0^\mathcal{S} - \lambda_0^{\mathcal{S} + P_k} $ is the difference between the $\mathcal{S}$ lowest eigenvalue and lowest eigenvalue obtained by including the operator $ P_k $, normalized by the largest such difference $(\Delta \lambda_0)_\text{max}$. Note that one should not interpret these eigenvalue decreases as representative of operator relevances in when reconstructing with the full basis, i.e., when reconstructing $H_\text{app}$ as in the main body of the paper. This is because the contributions to the lowest eigenvalue from each operator do not decompose additively in this manner. If a small number of operators from $\mathcal{P}$ are added to the basis $\mathcal{S}$, this decomposition holds approximately: 
for instance, in the projected Fermi sea a typical approximate decomposition has $\lambda_0^{\mathcal{S} + P_1}+ \lambda_0^{\mathcal{S} + P_2} \approx 5.80\mathrm{e}{-5}  $ and $ \lambda^{\mathcal{S} + P_1 + P_2} \approx 5.63\mathrm{e}{-4} $ for $P_1 = \mathcal{O}_{23}$ and $P_2 = \mathcal{O}_{35}$ as labelled in Fig.~\ref{fig:fsea-perms-evals}.

\nocite{*}

\bibliography{main}

\end{document}